\documentclass[preprint,12pt]{elsarticle}

\usepackage{amssymb}
\usepackage{subfigure}

\usepackage{lineno}
\usepackage{booktabs}
\usepackage{threeparttable}

\usepackage[usenames]{color}
\usepackage{algorithmic}
\usepackage{algorithm}
\usepackage{rotating}

\makeatletter
\def\ps@pprintTitle{%
  \let\@oddhead\@empty
  \let\@evenhead\@empty
  \let\@oddfoot\@empty
  \let\@evenfoot\@oddfoot
}
\makeatother

\begin{document}
\begin{frontmatter}



\title{Path Relinking for Bandwidth Coloring Problem}

\author[HUST,Angers]{Xiangjing Lai}
\ead{laixiangjing@gmail.com}
\author[HUST]{Zhipeng L\"u} 
\ead{zhipeng.lui@gmail.com}
\author[Angers]{Jin-Kao Hao}
\ead{hao@info.univ-angers.fr}
\author[USA]{Fred Glover}
\ead{glover@opttek.com}
\author[HUST]{Liping Xu}
\address[HUST]{SMART, School of Computer Science and Technology, \\Huazhong University of Science and Technology, 430074 Wuhan, P.R.China}
\address[Angers]{LERIA, Universit$\acute{e}$ d'Angers, 2 Boulevard Lavoisier, 49045 Angers, Cedex 01,  France}

\address[USA]{ECEE, Engineering \& Science, University of Colorado, Boulder, Colorado 80309, USA}

\begin{abstract}
A Path Relinking algorithm is proposed for the Bandwidth Coloring problem and the Bandwidth MultiColoring problem. It combines a population based relinking method and a tabu search based local search procedure. The proposed algorithm is assessed on two sets of 66 benchmark instances commonly used in the literature. Computational results demonstrate that the proposed algorithm is highly competitive in terms of both solution quality and efficiency compared to the best performing algorithms in the literature. Specifically, it improves the previous best known results for 15 out of 66 instances, while matching the previous best known results for 47 cases. Some key elements of the proposed algorithm are investigated.
\\
\noindent \emph{Keywords}:  Bandwidth Coloring, Path Relinking, Tabu Search, Heuristics, Frequency Assignment.
\end{abstract}
\end{frontmatter}


\section{Introduction}
\label{Intro}

Given an undirected graph $G = (V, E)$ with a set $V$ of vertices and an edge set $E$, the bandwidth coloring problem (BCP) is to assign a color $c_{i}$ ($1\le c_{i} \le k $ ) to each vertex  $i$ ($1\le i \le n$ ) such that for each edge $e(i,j)$ the difference between the colors of vertices $i$ and $j$ must be larger than or equal to the associated edge weight $d(i,j)$, i.e., $|c_{i} -c_{j} | \ge d(i,j)$. The objective of BCP is to minimize the number of the colors used, $k$. It is obvious that BCP is a generalization of the classical vertex coloring problem (VCP),which results in the case where $d(i,j)=1$ for all distinct pairs $(i,j)$.

The bandwidth multicoloring problem (BMCP) is a generalization of BCP, where each vertex  $i$ is associated with a positive integer $w(i)$ and each edge $e(i,j)$  is associated with an edge weight $d(i,j)$. BMCP aims to assign  $w(i)$ distinct colors from 1 to $k$ to each vertex $i$ ($1\le i \le n$), such that for each edge $e(i,j)$ the difference between the colors of vertices $i$ and $j$ must be at least the associated edge weight $d(i,j)$. Moreover, for $i=j$, the difference between any two colors of vertex $i$ must be at least $d(i,i)$, which is the weight of the loop edge of vertex $i$. Like BCP, BMCP aims to minimize the number of the colors used, $k$. One can observe that BCP is a special case of BMCP with $w(i)=1$ for all vertices.

In addition to their theoretical significance as NP-hard problems, BCP and BMCP have a number of relevant real-world applications. For example, the fixed spectrum frequency assignment problem (FS-FAP) \cite{Montemanni,Smith} can be viewed as a simple variant of BCP and the minimum span frequency assignment problem (MS-FAP) is equivalent to BMCP considered here \cite{Malaguti}. In addition, some other variants of BMCP have also been used to model the corresponding frequency assignment problems in the literature \cite{Ghosh,Wang} (see \cite{fap} for more details about the frequency assignment problems).

As mentioned in \cite{Dorne,Lim1,Malaguti}, BMCP can be converted into BCP by splitting each vertex $i$ into a clique with $w(i)$ vertices, where each edge of the clique has an edge weight $d(i,i)$ that corresponds to the weight of the loop edge of vertex $i$ in the original graph (see \cite{Dorne} for an example). The resulting new graph has $\sum_{i=1}^{n}w(i)$  vertices. Thus, any approach for BCP can be applied to BMCP directly. For this reason, we focus on solving BCP in this paper.

A large number of solution approaches have been reported in the literature. In 1998, Dorne and Hao proposed a tabu search algorithm for the T-coloring problem and the set T-coloring problem which are a generalization of BCP and BMCP \cite{Dorne}. In 2002, Phan and Skiena proposed a general heuristic \cite{Phan}, called Discropt for solving the vertex coloring problem and BCP, and Prestwich proposed a hybrid algorithm that combines a local search with constraint propagation for solving BCP and BMCP, called FCNS \cite{Prestwich1}. Subsequently, an extended version of FCNS was developed by the same author \cite{Prestwich2}. In \cite{Lim1,Lim2}, Lim et al. developed two hybrid algorithms for solving VCP and its generalizations. In \cite{Chi}, Chiarandini et al. investigate several stochastic local search algorithms for the set T-coloring problem. In 2008, Malaguti and Toth reported an effective evolutionary approach for BCP and BMCP \cite{Malaguti}. Recently, Mart\'{\i} et al. developed several heuristic approaches for BCP using memory structures in both constructive and improvement methods \cite{Marti}. In a very recent work \cite{Lai}, Lai and L\"{u} developed a multistart iterated tabu search (MITS) algorithm for BCP and BMCP. In addition, some other heuristic algorithms were also proposed in the literature to solve MS-FAP which is equivalent to BMCP, such as tabu search \cite{Costa,Hao1}, genetic algorithms \cite{Velanzuela}, and constraint programming approaches \cite{Walser}.

Recently, Path Relinking (PR) \cite{Glover1,Glover2} has attracted special attention in the community of combinatorial optimization, and shows outstanding performances in solving a number of difficult problems, such as unconstrained binary quadratic optimization \cite{Wangyang},  multiple-level warehouse layout \cite{Zhang}, and  flow shop sequencing \cite{Reeves}. In this paper, we devise a new PR algorithm for the BCP and BMCP problems, which integrates a tabu search (TS) algorithm (for local optimization)  with the population based PR framework. The proposed PR algorithm is assessed on two sets of 66 benchmark instances commonly used in the literature and shows remarkable performances compared to the current best solution methods.

The rest of this paper is organized as follows. In Section \ref{PR} , we describe in detail the PR algorithm proposed in this paper. In Section \ref{results}, we show our computational results compare them with several best performing algorithms from the literature. In Section \ref{discussion}, we investigate some key ingredients of the PR algorithm, before concluding the paper in Section \ref{Discussion}.

\section{Path Relinking Algorithm}
\label{PR}
The PR algorithm presented in this paper is a hybrid population algorithm that combines path relinking and local search to achieve a desirable tradeoff between intensification and diversification. The effectiveness of the PR algorithm depends mainly on three components: the PR population scheme, the path relinking procedure, and the local search method.  We explain in this section the ingredients of our proposed PR algorithm designed for BCP.

\subsection{Search Space and Objective Function for $k$-BCP}
\label{sub_search_space}

BCP can be considered from the point of view of constraint satisfaction by solving a series of $k$-BCP problems aiming at searching for a $k$-coloring ($k$ being fixed) that satisfies all edge constraints. Starting from a large enough initial $k$, our algorithm seeks to solve the $k$-BCP problem, i.e. to find a legal $k$-coloring. A $k$-coloring is legal if all the edge constraints of BCP are satisfied and is illegal, otherwise. As soon as the $k$-BCP problem is solved for the current $k$ value, we set $k$ to $k-1$ and solve again the new $k$-BCP problem. The process is repeated until no legal $k$-coloring can be found. Therefore, the presented PR algorithm only considers the $k$-BCP problem.

In general, a combinatorial optimization problem can be represented as a 2-tuple $(S,f)$, where $S$ represents the search space and $f$ represents the objective function defined on $S$. As stated in \cite{Lai}, for the $k$-BCP problem, the search space $S$ can be defined as the set of all possible $k$-colorings, including legal and illegal $k$-colorings.  It should be noted that the same search space is also used for the studies of vertex coloring problem in for instance \cite{Lu,Wu}.

Moreover, for a $k$-coloring, we define the objective function of the $k$-BCP problem as the summation of all constraint violations induced by the $k$-coloring. Specifically, let $s$ be a $k$-coloring, the objective function $f(s)$ used in this study is written as:
 \begin{equation}
\centering
f(s)=\sum_{e(i,j)\in{E}}max\{{0,d(i,j)-|c_i-c_j|}\}
\end{equation}
where $d(i,j)$ is the edge weight for edge $e(i, j)$, and $c_i$  and $c_j$  respectively represent colors of vertices  $i$ and $j$. Therefore, a solution  $s$ with  $f(s)=0$ corresponds to a legal $k$-coloring.

\subsection{Main Framework }
\label{subsec_Main_Scheme}

As mentioned above, our PR algorithm falls into the class of population algorithms and consists of four main components: population initialization, local search method, path relinking procedure, and population update.

Algorithm \ref{Algo_PR} describes in detail the framework of our PR algorithm. In the algorithm, $s^*$ and $s^w$ respectively represent the best solution found so far and the worst solution in the population in terms of objective function value, $PairSet$ is the set of solution pairs $(s_i, s_j)$ and is initially composed of all the possible solution pairs $(s_i, s_j)$ in the population. The PR algorithm starts with an initial population $P$ (line 4) which includes $p$ different solutions, where each of them is randomly generated and then optimized by the tabu search procedure to reach a local optimum. Subsequently, the algorithm enters a while loop (lines 11 to 25), and at each iteration a solution pair is randomly selected from $PairSet$, and then the path relinking procedure (line 14) and tabu search procedure (line 15) are applied to generate two new offspring solutions. More specifically, for the selected solution pair $(s_i, s_j)$, two offspring solutions $s^{'}$ and $s^{''}$ are generated by building paths from the initial solution $s_i$ to the target solution $s_j$ as well as from $s_j$ to $s_i$, and then $s^{'}$ and $s^{''}$ are respectively optimized by the tabu search procedure. After that, a population update criterion is used to decide whether the obtained solutions should be inserted into the population (lines 19 to 20). The $PairSet$ is updated as follows. First, the solution pair $(s_i, s_j)$ is removed after it has been chosen from $PairSet$ for the path relinking procedure (line 13). Second, after the population update, once a reference solution $s^{'}$ replaces the worst solution $s^w$ in the population, all the solution pairs containing $s^w$ are removed from $PairSet$ and all the solution pairs that can be generated by combining $s^{'}$ with other solutions in the population are added into $PairSet$ (lines 21 to 22).  The while loop ends when $PairSet$ becomes empty, then the population is recreated (lines 4 to 7) and the above while loop is repeated if the stopping condition is not satisfied.

There are several stopping criteria that can be employed for the PR algorithm, such as the maximum number of iterations, the maximum number of iterations during which the best solution cannot be improved, the maximum timeout limit, and so on. In this study, our PR algorithm stops when a legal solution ($k$-coloring) is found or the timeout limit is reached.

The PR algorithm presented in this paper has several specific features. First, each solution pair $(s_i, s_j)$ is used to generate two paths by the path relinking procedure: one is from $s_i$ to $s_j$, the other is from $s_j$ to $s_i$. Therefore, each selected solution pair will generate two new offspring solutions by the path relinking procedure. Second, when $PairSet$ is empty, we rebuild the population while retaining the best solution $s^{*}$ in the new population (lines 4 to 7).

In the following subsections, we describe in detail each main component of our PR algorithm.

\subsection{Population Initialization}
\label{subsec_Initialization_population}

From scratch, an initial population is constructed as follows.  We first generate $3p$ random solutions, where each variable (or vertex) of each solution is assigned a random color from 1 to $k$. Then, for each generated solution,  the tabu search method (see Section \ref{subsec_TS}) is used to optimize it to a local optimum. Finally, we choose the first $p$ best solutions to form the initial population.

\begin{algorithm}[h]
\begin{small}
 \caption{Pseudo-code of our PR algorithm for BCP} \label{Algo_PR}
 \begin{algorithmic}[1]
   \STATE \sf \textbf{Input}: Problem instance $I$, the number of colors ($k$), the size of population ($p$)
   \STATE \textbf{Output}: the best $k$-coloring $s^{*}$ found and $f(s^{*})$
   \REPEAT
   \STATE $P=\{s^{1},\ldots,s^{p}\}$ $\leftarrow$
   Population\_Initialization($I,k$) \ \ \ \ \ \ \ \ \ \ /$*$ Section \ref{subsec_Initialization_population} $*$/
   \IF{it is not in the first loop}
   \STATE $P \leftarrow P \cup \{s^{*}\} \setminus \{s^{w}\}$
   \ENDIF
   \STATE $s^{*}=arg \ min \{f(s^{i})|i=1,\ldots,p\}$
   \STATE $s^{w}=arg \ max \{f(s^{i})|i=1,\ldots,p\}$
   \STATE $PairSet \leftarrow \{(s^i,s^j)\mid 1\leq i < j \leq p \}$
   \WHILE{$PairSet\neq \varnothing$}
        \STATE Randomly pick a solution pair $(s^i,s^j)\in PairSet$
        \STATE $PairSet \leftarrow PairSet \setminus \{(s^i,s^j)\}$
        \STATE $s^{'} \leftarrow Path Relinking(s^i,s^j)$,
        $s^{''} \leftarrow Path Relinking(s^j,s^i)$/$*$ Section \ref{subsec_Path_Relinking_operator} $*$/
        \STATE $s^{'} \leftarrow Local Search (s^{'})$  \ \ \ \ \ \ \ \ \ \ \ \ \ \ \ \ \ \ \ \ \ \ \ \ \ \ \ \ \ \ \ \ \ \ \ \ \ \ \  \ \ \ \ \ \ \ /$*$ Section \ref{subsec_TS} $*$/
        \IF{$f(s^{'}) < f(s^{*})$}
              \STATE $s^{*} = s^{'}$, $f(s^{*})= f(s^{'}) $
        \ENDIF
        \IF{the update criterion is satisfied}
        \STATE $P \leftarrow P \cup \{s^{'}\} \setminus \{s^{w}\}$
        \STATE $PairSet \leftarrow PairSet \setminus \{(s^w,s^k)\mid s^k \in P\}$
        \STATE $PairSet \leftarrow PairSet \cup \{(s^{'},s^k)\mid s^k \in P\}$
        \ENDIF  \ \ \ \ \ \ \ \ \ \ \ \ \ \ \ \ \ \ \ \ \ \ \ \ \ \ \ \ \ \ \ \ \ \ \ \ \ \ \ \ \ \ \ \ \ \ \ \ \ \ \ \ \ \ \ \ \ \ \ \ \ \ \ \ \ \ \ /$*$ Section \ref{subsec_pool_updating} $*$/
   \STATE Perform the same steps from line 15 to line 23 for $s^{''}$
   \ENDWHILE
   \UNTIL{a stop criterion is met}
 \end{algorithmic}
 \end{small}
\end{algorithm}

\subsection{Local Optimization with Tabu Search}
\label{subsec_TS}

In \cite{Lai}, an effective tabu search (TS) algorithm is presented for the BCP and BMCP problems. In this work, we use this TS algorithm as our local optimization procedure. The main characteristics are as follows. The TS procedure operates in the same search space and objective function as defined in  Section \ref{sub_search_space}. From a given solution (i.e., an illegal $k$-coloring), it iteratively improves the current solution by following the ``critical one-move'' neighborhood. As such, a neighboring solution of the current $k$-coloring is obtained by changing the color of a conflicting vertex $u$ from its original color $c_i$ to another color $c_j$ ($c_i\neq c_j$)  (denoted by the move $<u,c_i,c_j>$). Here, a vertex is considered to be conflicting if at least one of the distance constraints associated with this vertex is violated. Thus, it can easily be deduced that for a $k$-coloring $s$ with cost $f(s)$, the size of this neighborhood is bounded by $O(f(s)\times k)$. At each iteration of TS, a best authorized neighboring solution is chosen to replace the current solution and the corresponding move is recorded to the tabu list to prevent the reverse move from being selected during the next $tl$ iterations ($tl$ is the tabu tenure). For a detailed description of the TS optimization procedure, the reader is refereed to  \cite{Lai}.

\subsection{The Relinking Procedure}
\label{subsec_Path_Relinking_operator}
The relinking method is the core component of our PR algorithm, and its goal is to generate new solutions by creating paths between two high-quality solutions. The relinking method includes two main operations. The first one is to construct a path that connects two parent solutions, where the parent solutions located at the beginning and the end of the path are respectively called the initiating solution and the guiding solution, while the others are called intermediate solutions.  The other operation is to choose one solution as the reference solution from the constructed path and apply the TS procedure to improve its quality.

To build a path between the initial and the guiding solutions, there are generally two simple strategies that can be used, i.e., the random strategy (denoted by PR1) and the greedy strategy (denoted by PR2). In this study, we employ PR1, since PR1 is superior to PR2 within the present framework according to our preliminary experiments.

The pseudo-code of these two strategies are described in Algorithms \ref{Algo_RK} and \ref{Algo_RK2}, where some important definitions are given as follows:
\begin{itemize}
  \item $NC$ : the set of variable indices that have different values in two solutions.
  \item $\Delta_t$ : the change in the objective function value when the current solution $s_i$ is changed by modifying its $t$th variable to transit from $s_i$ to $s_j$.
  \item $PathSet$ : a variable set that stores the variables modified at each step throughout the transiting from $s_i$ to $s_j$.
  \item $Q_{n \times k}$ : a memory structure for updating $\Delta_t$ in a fast manner.
\end{itemize}

\begin{algorithm}[h]
\begin{small}
 \caption{Pseudo-code of constructing a path for PR1} \label{Algo_RK}
 \begin{algorithmic}[1]
   \STATE \sf \textbf{Input}: A pair of solutions $(s_i,s_j)$
   \STATE \textbf{Output}: Path solutions $s(0),s(1),\dots, s(r)$ from $s_i$ to $s_j$
   \STATE $NC = \{l|s_i^{l}\neq s_j^{l},l=1,2,\dots,n \}$
   \STATE $PathSet= \varnothing$, $r=|NC|-1$, $s(0)=s_i$
   \STATE Initialize $Q_{n \times k}$
   \FOR{$m=\{1,\ldots,r\}$}
       \STATE Choose a $t \in NC$ at random
       \STATE $\Delta_t$ = $Q[t][s_j^{t}]-Q[t][s_i^{t}]$
       \STATE $PathSet \leftarrow PathSet \cup \{t\} $
       \STATE $s(m)= \{ s^{u}: s^{u} = s_j^{u},u\in PathSet; s^{u} = s_i^{u},u \in N \setminus PathSet;$ \}
       \STATE $f(s(m)) = f(s(m-1)) + \Delta_t $
       \STATE $NC \leftarrow NC\setminus \{t\}$
       \STATE Update $Q_{n \times k}$
   \ENDFOR

 \end{algorithmic}
 \end{small}
\end{algorithm}

\begin{algorithm}[h]
\begin{small}
 \caption{Pseudo-code of constructing a path for PR2} \label{Algo_RK2}
 \begin{algorithmic}[1]
   \STATE \sf \textbf{Input}: A pair of solutions $(s_i,s_j)$
   \STATE \textbf{Output}: Path solutions $s(0),s(1),\dots, s(r)$ from $s_i$ to $s_j$
   \STATE $NC = \{l|s_i^{l}\neq s_j^{l},l=1,2,\dots,n \}$
   \STATE $PathSet= \varnothing$, $r=|NC|-1$, $s(0)=s_i$
   \STATE Initialize $Q_{n \times k}$
   \FOR{$m=\{1,\ldots,r\}$}
       \STATE $\Delta_{min}$ = $min\{\Delta_{t}|\Delta_{t} = Q[t][s_j^{t}]-Q[t][s_i^{t}], t \in NC\}$
       \STATE Choose a $t \in NC$ with $ \Delta_{t} = \Delta_{min} $
       \STATE $PathSet \leftarrow PathSet \cup \{t\} $
       \STATE $s(m)= \{ s^{u}: s^{u} = s_j^{u},u\in PathSet; s^{u} = s_i^{u},u \in N \setminus PathSet;$ \}
       \STATE $f(s(m)) = f(s(m-1)) + \Delta_{min} $
       \STATE $NC \leftarrow NC\setminus \{t\}$
       \STATE Update $Q_{n \times k}$
   \ENDFOR

 \end{algorithmic}
 \end{small}
\end{algorithm}

In the path relinking procedures described Algorithms \ref{Algo_RK} and \ref{Algo_RK2}, a solution sequence (i.e., a path) of length of $r+1$: ($s(0)$, $s(1)$, $s(2)$ $\dots$ $s(r)$) is generated in a step by step way by starting from $s(0)$, where $s(m)$ differs from $s(m-1)$ by the value of only one variable, $m  = 1, 2 \dots r$, and $r+1$  represents the size of the initial $NC$.  In addition, $s(0)$ and $s(r)$ correspond respectively to the guiding solution $s_i$ and the target solution $s_j$, while the other solutions are intermediate solutions.

The only difference between PR1 and PR2 lies in the way to select the variable $t$ from $NC$ for generating the next solution on the path. In PR1, based on the current solution $s(i)$ a random variable $t$ ($t\in NC$) is chosen to generate the next solution $s(i+1)$, $i= 0 ,1, \dots, r-1$ , whereas in PR2 a variable $t$ producing the best $\Delta_{t}$ value is chosen.

Note that $\Delta_t$ can be rapidly calculated at each step of the path relinking with the help of the dedicated memory structure $Q_{n \times k}$ \cite{Lai}, and this matrix is only initialized at the beginning of each relinking procedure and updated subsequently in a very fast manner during the relinking process.

After the creation of a path, we choose one solution on this path such that the chosen solution is far enough from the initiating and target solutions and has a good objective function value \cite{Wangyang}.  Specifically, we construct a candidate solution list (CLS) that consists of the path solutions having a distance of at least $\xi\cdot|NC|$ (where $\xi$ is a value between 0 and 1.0 and $|NC|$ is the Hamming distance between the initiating and guiding solutions) from both the initiating and guiding solutions, then the solution having the best objective value in CSL is chosen as the reference solution which is further improved by tabu search.

\subsection{Population Updating}
\label{subsec_pool_updating}
To determine whether a reference solution should be inserted into the population and which solution in the population should be replaced, we use the following updating criterion. The reference solution replaces the worst solution in the population if it is not too close to any solution in the population and it is better than the worst solution in the population. Two solutions are considered to be too close if the number of components that have different values is smaller than $0.1\times n$, where $n$ is the number of vertices in the graph.

\section{Experimental Results and Comparisons}
\label{results}
\subsection{Instances and Experimental Protocol}

Two sets of benchmarks are considered in our experiments. The first set (denotaed by BCP) is composed of 33 instances and available in \cite{Trick}. The second set (denotaed by BMCP)  is composed of 33 larger instances, transformed from the BMCP instances.

Our algorithm was programmed in C and run on a cluster with 2.8 GHz CPU and 4Gb RAM. Table \ref{Parameter_Settings} gives the descriptions and settings of the parameters used in our PR algorithm.  For the first and second sets of benchmark instances, the timeout limits are set to 2 and 4 hours, respectively. For the majority of the tested instances the actual time to reach a legal $k$-coloring is generally much less than these time limits. Given the stochastic nature of our PR algorithm, each instance with the given value of $k$ is independently solved  20 times.

\renewcommand{\baselinestretch}{1.0}\large\normalsize
\begin{table}\centering
\begin{scriptsize}
\caption{Settings of important parameters}
\label{Parameter_Settings}
\begin{tabular}{p{1.4cm}p{1.2cm}p{4.9cm}p{0.01cm}p{1.3cm}p{0.01cm}}
\hline
Parameters &Section& Description & & Values  & \\
\hline
$p$       & \ref{subsec_Main_Scheme}   & population size                            & & 20            \\
$\alpha$  & \ref{subsec_TS}            & depth of TS                         & & $10^{4}$               & \\
$\xi$     & \ref{subsec_Path_Relinking_operator}  & distance parameter            & & $0.35$                 &\\

\hline
\end{tabular}
\end{scriptsize}
\end{table}
\renewcommand{\baselinestretch}{1.0}\large\normalsize

\subsection{Computational Results and Comparison on the BCP Instances}
\label{bcp}

\renewcommand{\baselinestretch}{1.0}\huge\normalsize
\begin{table}\centering
\caption{Comparison of the PR algorithm with other algorithms on BCP instances}  \label{results_bcp_instances}
\begin{scriptsize}
\begin{tabular}{p{1.35cm}|p{0.3cm}|p{0.3cm}|p{0.3cm}|p{0.3cm}|p{0.3cm}p{0.5cm}|p{0.3cm}p{0.6cm}|p{0.3cm}p{0.5cm}p{1.0cm}p{0.8cm}p{0.01cm}}
  \hline
  &  & \multicolumn{1}{c}{\cite{Phan}}&\multicolumn{1}{c}{\cite{Lim2}} & \multicolumn{1}{c}{\cite{Prestwich2}}& \multicolumn{2}{c}{\cite{Malaguti}}&\multicolumn{2}{c}{MITS \cite{Lai}}& \multicolumn{4}{c}{PR Algorithm} &\\
  \cline{3-13}
  \centering{{Instance}}&\centering{{$k^{*}$}}&\centering {$k$} &\centering{$k$}&\centering{$k$} & \centering{$k$} & \centering{T(s)}& \centering{$k$}&\centering{$T_{ave}$}& \centering{$k$}&\centering{ suc } & \centering{$T_{ave}$}(s)& \centering{$k-k^*$}&\\
  \hline
  \centering{GEOM20}& \centering{20}& \centering{20}&\centering{21}&\centering{21}&\centering{21}& \centering{0}&\centering{21}&\centering{0}&\centering{21}&\centering{20/20}& \centering{0}& \centering{1}& \\
  \centering{GEOM20a}& \centering{20}& \centering{20}&\centering{22}&\centering{20}&\centering{20}& \centering{0}&\centering{20}&\centering{0}&\centering{20}&\centering{20/20}& \centering{0}& \centering{0}&\\
  \centering{GEOM20b}& \centering{13}& \centering{13}&\centering{14}&\centering{13}&\centering{13}& \centering{0}&\centering{13}&\centering{0}&\centering{13}&\centering{20/20}& \centering {0}& \centering{0}&\\
  \centering{GEOM30}& \centering{27}& \centering{27}&\centering{29}&\centering{28}&\centering{28}& \centering{0}&\centering{28}&\centering{0}&\centering{28}&\centering{20/20}& \centering{0}& \centering{1}&\\
  \centering{GEOM30a}& \centering{27}& \centering{27}&\centering{32}&\centering{27}&\centering{27}& \centering{0}&\centering{27}&\centering{0}&\centering{27}&\centering{20/20}& \centering{0}& \centering{0}&\\
  \centering{GEOM30b}& \centering{26}& \centering{26}&\centering{26}&\centering{26}&\centering{26}& \centering{0}&\centering{26}&\centering{0}&\centering{26}&\centering{20/20}& \centering{0}& \centering{0}&\\
  \centering{GEOM40}& \centering{27}& \centering{27}&\centering{28}&\centering{28}&\centering{28}& \centering{0}&\centering{28}&\centering{0}&\centering{28}&\centering{20/20}& \centering{0}& \centering{1}&\\
  \centering{GEOM40a}& \centering{37}& \centering{38}&\centering{38}&\centering{37}&\centering{37}& \centering{0}&\centering{37}&\centering{0}&\centering{37}&\centering{20/20}& \centering{0}& \centering{0}&\\
  \centering{GEOM40b}& \centering{33}& \centering{36}&\centering{34}&\centering{33}&\centering{33}& \centering{0}&\centering{33}&\centering{0}&\centering{33}&\centering{20/20}& \centering{0}& \centering{0}&\\
  \centering{GEOM50}& \centering{28}& \centering{29}&\centering{28}&\centering{28}&\centering{28}& \centering{0}&\centering{28}&\centering{0}&\centering{28}&\centering{20/20}& \centering{0}& \centering{0}&\\
  \centering{GEOM50a}& \centering{50}& \centering{54}&\centering{52}&\centering{50}&\centering{50}& \centering{0}&\centering{50}&\centering{0}&\centering{50}&\centering{20/20}& \centering{0}& \centering{0}&\\
  \centering{GEOM50b}& \centering{35}& \centering{40}&\centering{38}&\centering{35}&\centering{35}& \centering{0}&\centering{35}&\centering{3}&\centering{35}&\centering{20/20}& \centering{1}& \centering{0}&\\
  \centering{GEOM60}& \centering{33}& \centering{34}&\centering{34}&\centering{33}&\centering{33}& \centering{0}&\centering{33}&\centering{0}&\centering{33}&\centering{20/20}& \centering{0}& \centering{0}&\\
  \centering{GEOM60a}& \centering{50}& \centering{54}&\centering{53}&\centering{50}&\centering{50}& \centering{0}&\centering{50}&\centering{1}&\centering{50}&\centering{20/20}& \centering{0}& \centering{0}&\\
  \centering{GEOM60b}& \centering{41}& \centering{47}&\centering{46}&\centering{43}&\centering{41}& \centering{29}&\centering{41}&\centering{277}&\centering{41}&\centering{20/20}& \centering{105}& \centering{0}&\\
  \centering{GEOM70}& \centering{38}& \centering{40}&\centering{38}&\centering{38}&\centering{38}& \centering{0}&\centering{38}&\centering{0}&\centering{38}&\centering{20/20}& \centering{0}& \centering{0}&\\
  \centering{GEOM70a}& \centering{61}& \centering{64}&\centering{63}&\centering{62}&\centering{61}& \centering{12}&\centering{61}&\centering{45}&\centering{61}&\centering{20/20}& \centering{47}& \centering{0}&\\
 \centering{GEOM70b}& \centering{47}& \centering{54}&\centering{54}&\centering{48}&\centering{48}& \centering{52}&\centering{47}&\centering{8685}&\centering{47}&\centering{12/20}& \centering{6678}& \centering{0}&\\
 \centering{GEOM80}& \centering{41}& \centering{44}&\centering{42}&\centering{41}&\centering{41}& \centering{0}&\centering{41}&\centering{0}&\centering{41}&\centering{20/20}& \centering{0}& \centering{0}&\\
 \centering{GEOM80a}& \centering{63}& \centering{69}&\centering{66}&\centering{63}&\centering{63}& \centering{150}&\centering{63}&\centering{21}&\centering{63}&\centering{20/20}& \centering{12}& \centering{0}&\\
 \centering{GEOM80b}& \centering{60}& \centering{70}&\centering{65}&\centering{61}&\centering{60}& \centering{145}&\centering{60}&\centering{322}&\centering{60}&\centering{20/20}& \centering{191}& \centering{0}&\\
 \centering{GEOM90}& \centering{46}& \centering{48}&\centering{46}&\centering{46}&\centering{46}& \centering{0}&\centering{46}&\centering{0}&\centering{46}&\centering{20/20}& \centering{0}& \centering{0}&\\
 \centering{GEOM90a}& \centering{63}& \centering{74}&\centering{69}&\centering{64}&\centering{63}& \centering{150}&\centering{63}&\centering{230}&\centering{63}&\centering{20/20}& \centering{191}& \centering{0}&\\
 \centering{GEOM90b}& \centering{69}& \centering{83}&\centering{77}&\centering{72}&\centering{70}& \centering{1031}&\centering{69}&\centering{20144}&\centering{69}&\centering{5/20}& \centering{23560}& \centering{0}&\\
 \centering{GEOM100}& \centering{50}& \centering{55}&\centering{51}&\centering{50}&\centering{50}& \centering{2}&\centering{50}&\centering{2}&\centering{50}&\centering{20/20}& \centering{2}& \centering{0}&\\
 \centering{GEOM100a}& \centering{67}& \centering{84}&\centering{76}&\centering{68}&\centering{68}& \centering{273}&\centering{67}&\centering{11407}&\centering{67}&\centering{14/20}& \centering{5556}& \centering{0}&\\
 \centering{GEOM100b}& \centering{72}& \centering{87}&\centering{83}&\centering{73}&\centering{73}& \centering{597}&\centering{72}&\centering{24561}&\centering{72}&\centering{3/20}& \centering{41832}& \centering{0}&\\
 \centering{GEOM110}& \centering{50}& \centering{59}&\centering{53}&\centering{50}&\centering{50}& \centering{3}&\centering{50}&\centering{2}&\centering{50}&\centering{20/20}& \centering{5}& \centering{0}&\\

 \centering{GEOM110a}& \centering{72}& \centering{88}&\centering{82}&\centering{73}&\centering{72}& \centering{171}&\centering{72}&\centering{1529}&\centering{\textbf{71}}&\centering{13/20}& \centering{5140}& \centering{-1}&\\
 \centering{}& \centering{}& \centering{}&\centering{}&\centering{}&\centering{}& \centering{}&\centering{}&\centering{}&\centering{72}&\centering{20/20}& \centering{552}& \centering{0}&\\
 \centering{GEOM110b}& \centering{78}& \centering{87}&\centering{88}&\centering{79}&\centering{78}& \centering{676}&\centering{78}&\centering{24416}&\centering{78}&\centering{6/20}& \centering{18136}& \centering{0}&\\
 \centering{GEOM120}& \centering{59}& \centering{67}&\centering{62}&\centering{60}&\centering{59}& \centering{0}&\centering{59}&\centering{1}&\centering{59}&\centering{20/20}& \centering{2}& \centering{0}&\\
 \centering{GEOM120a}& \centering{82}& \centering{101}&\centering{92}&\centering{84}&\centering{84}& \centering{614}&\centering{82}&\centering{34176}&\centering{82}&\centering{2/20}& \centering{62876}& \centering{0}&\\
  \centering{GEOM120b}& \centering{84}& \centering{103}&\centering{98}&\centering{86}&\centering{84}& \centering{857}&\centering{84}&\centering{inf}&\centering{85}&\centering{2/20}& \centering{66301}& \centering{1}&\\
 \hline
\end{tabular}
\end{scriptsize}
\end{table}
\renewcommand{\baselinestretch}{1.0}\large\normalsize

Our first experiment aims to evaluate the PR algorithm on the set of 33 BCP instances with up to 120 vertices. For each instance, only those values of $k$ which are close to the previous best known results are tested. Table \ref{results_bcp_instances} summarizes the computational statistics and includes the results of 5 reference algorithms from the literature. Column 2 gives the previous best known results ($k^{*}$). Columns 3 to 9 respectively show the results of the 5 reference algorithms, namely the Discropt heuristic \cite{Phan}, a hybrid algorithm \cite{Lim2}, the FCNS algorithm \cite{Prestwich2}, an evolutionary algorithm \cite{Malaguti}, and the MITS algorithm \cite{Lai}. The computational statistics of our PR algorithm is reported in columns 10 to 13, including the best $k$ value achieved ($k$), the success rate ($suc$) with the associated $k$ value, the average time per hit of legal $k$-coloring ($T_{ave}$) that is obtained by dividing the total time by the number of successful times, and the difference between the given value of $k$ and the previous best known result ($k-k^{*}$), where the improved results are indicated in bold and ``inf'' means that the corresponding algorithm fails to find a legal $k$-coloring.

From Table \ref{results_bcp_instances}, one observes that our PR algorithm improves the previous best-known result on one instance and obtains worse results on 4 instances. For the other 28 instances, our algorithm matches the previous best known results. Nevertheless, like other 4 reference algorithms, our PR algorithm is not able to reach the previously reported best known results for the three small instances, namely GEOM20, GEOM30, and GEOM40. Notice that the best known results for these three small graphs were reported only by one simple algorithm \cite{Phan} and never confirmed by other more advanced methods.

In addition, compared with the best results of the first four reference algorithms, the present PR algorithm is able to obtain better results on at least 6 instances though with a longer computing time, and worse result on one instance. Relative to these instances containing no more than 120 vertices, it should be noted that both the MITS algorithm and the PR algorithm, which share the same local search algorithm, obtain the same best results except for the instance named GEOM110a. For this instance, however, the performance of the methods differed substantially. The MITS algorithm failed to obtain the current best result with a very large computational effort, while the present PR algorithm is able to find the current best result with an average computational time of 5140 seconds. Given that our PR algorithm solves $k$-BCP, it would be difficult to make a fair comparison of the computational time with reference algorithms. The CPU times of the first 4 reference algorithms are given only for indicative purposes.

Furthermore, the success rate of our PR algorithm is larger than 12/20 except for 5 difficult instances, demonstrating the robustness of our algorithm.

It should be noticed that our reported computing times have been scaled with respect to the performance obtained with a common benchmark program (dfmax) and a benchmark instance (R500.5) which is available in \cite{Trick} (see \cite{Lai} for more details). Therefore, according to the spirit of \cite{Malaguti},  the computing times are in some sense comparable with those obtained with these reference algorithms.

\subsection{Comparison of the PR algorithm with Other Algorithms on the BMCP Instances}
\label{bmcp}

\renewcommand{\baselinestretch}{1.0}\huge\normalsize
\begin{table}\centering
\caption{ Comparison of the PR algorithm with other algorithms on BMCP instances} \label{results_bmcp_instances}
\begin{scriptsize}
\begin{threeparttable}
\begin{tabular}{p{1.4cm}|p{0.3cm}|p{0.3cm}|p{0.2cm}|p{0.2cm}|p{0.2cm}p{0.3cm}|p{0.3cm}p{0.5cm}|p{0.3cm}p{0.72cm}|p{0.4cm}p{0.4cm}p{1.0cm}p{0.01cm}p{0.01cm}}
\toprule

  &  & \multicolumn{1}{c}{\cite{Lim2}}& \multicolumn{1}{c}{\cite{Lim1}}& \multicolumn{1}{c}{\cite{Prestwich2}} & \multicolumn{2}{c}{OF-SW \cite{Chi}}  & \multicolumn{2}{c}{MA\cite{Malaguti}} &  \multicolumn{2}{c}{MITS \cite{Lai}} & \multicolumn{4}{c}{PR} & \\
  \cline{3-16}
  \centering{{Instance}} & \centering{{$k^{*}$}}& \centering {$k$} & \centering{$k$}&\centering{$k$} & \centering{$k$} & \centering{$T_{max}$} &\centering{$k$} & \centering{T(s)}& \centering{$k$} & \centering{$T_{ave}$} & \centering{$k_{best}$} & \centering{SR} & \centering{$T_{ave}$(s)} & &\\
  \hline
   \centering{GEOM20}& \centering{149}& \centering{149}&\centering{149}&\centering{149}&\centering{-}&\centering{-}&\centering{149}& \centering{18}&\centering{149}&\centering{2}&\centering{149}&\centering{20/20}& \centering{1}&\\
  \centering{GEOM20a}& \centering{169}& \centering{169}&\centering{169}&\centering{170}&\centering{-}&\centering{-}&\centering{169}& \centering{9}&\centering{169}&\centering{15}&\centering{169}&\centering{20/20}& \centering{7}& \\
  \centering{GEOM20b}& \centering{44}& \centering{44}&\centering{44}&\centering{44}&\centering{44}&\centering{30}&\centering{44}& \centering{5}&\centering{44}&\centering{0}&\centering{44}&\centering{20/20}& \centering {0}& \\
  \centering{GEOM30}& \centering{160}& \centering{160}&\centering{160}&\centering{160}&\centering{-}&\centering{-}&\centering{160}& \centering{1}&\centering{160}&\centering{0}&\centering{160}&\centering{20/20}& \centering{0}& \\
  \centering{GEOM30a}& \centering{209}& \centering{211}&\centering{209}&\centering{214}&\centering{209}&\centering{380}&\centering{210}& \centering{954}&\centering{209}&\centering{10}&\centering{209}&\centering{20/20}& \centering{26}& \\
  \centering{GEOM30b}& \centering{77}& \centering{77}&\centering{77}&\centering{77}&\centering{77}&\centering{80}&\centering{77}& \centering{0}&\centering{77}&\centering{0}&\centering{77}&\centering{20/20}& \centering{0}& \\
  \centering{GEOM40}& \centering{167}& \centering{167}&\centering{167}&\centering{167}&\centering{-}&\centering{-}&\centering{167}& \centering{20}&\centering{167}&\centering{0}&\centering{167}&\centering{20/20}& \centering{1}& \\
  \centering{GEOM40a}& \centering{213}& \centering{214}&\centering{213}&\centering{217}&\centering{214}&\centering{500}&\centering{214}& \centering{393}&\centering{213}&\centering{328}&\centering{213}&\centering{20/20}& \centering{133}& \\
  \centering{GEOM40b}& \centering{74}& \centering{76}&\centering{74}&\centering{74}&\centering{74}&\centering{140}&\centering{74}& \centering{1}&\centering{74}&\centering{2}&\centering{74}&\centering{20/20}& \centering{4}& \\
  \centering{GEOM50}& \centering{224}& \centering{224}&\centering{224}&\centering{224}&\centering{-}&\centering{-}&\centering{224}& \centering{1197}&\centering{224}&\centering{8}&\centering{224}&\centering{20/20}& \centering{2}& \\
  \centering{GEOM50a}& \centering{314}& \centering{326}&\centering{318}&\centering{323}&\centering{315}&\centering{1080}&\centering{316}& \centering{4675}&\centering{314}&\centering{40373}&\centering{314}&\centering{14/20}& \centering{10232}& \\
  \centering{GEOM50b}& \centering{83}& \centering{87}&\centering{87}&\centering{86}&\centering{84}&\centering{200}&\centering{83}& \centering{197}&\centering{83}&\centering{1200}&\centering{83}&\centering{20/20}& \centering{723}& \\
  \centering{GEOM60}& \centering{258}& \centering{258}&\centering{258}&\centering{258}&\centering{258}&\centering{710}&\centering{258}& \centering{139}&\centering{258}&\centering{19}&\centering{258}&\centering{20/20}& \centering{23}& \\
  \centering{GEOM60a}& \centering{356}& \centering{368}&\centering{358}&\centering{373}&\centering{356}&\centering{1420}&\centering{357}& \centering{8706}&\centering{356}&\centering{38570}&\centering{356}&\centering{18/20}& \centering{5741}& \\
  \centering{GEOM60b}& \centering{113}& \centering{119}&\centering{116}&\centering{116}&\centering{117}&\centering{300}&\centering{115}& \centering{460}&\centering{113}&\centering{104711}&\centering{113}&\centering{4/20}& \centering{63579}& \\
  \centering{GEOM70}& \centering{267}& \centering{279}&\centering{273}&\centering{277}&\centering{267}&\centering{1060}&\centering{272}& \centering{1413}&\centering{270}&\centering{7602}&\centering{270}&\centering{20/20}& \centering{561}& \\
  \centering{GEOM70a}& \centering{467}& \centering{478}&\centering{469}&\centering{482}&\centering{478}&\centering{1470}&\centering{473}& \centering{988}&\centering{467}&\centering{38759}&\centering{467}&\centering{20/20}& \centering{5212}& \\
 \centering{GEOM70b}& \centering{116}& \centering{124}&\centering{121}&\centering{119}&\centering{120}&\centering{380}&\centering{117}& \centering{897}&\centering{116}&\centering{213545}&\centering{116}&\centering{8/20}& \centering{26110}& \\
 \centering{GEOM80}& \centering{381}& \centering{394}&\centering{383}&\centering{398}&\centering{382}&\centering{1490}&\centering{388}& \centering{132}&\centering{381}&\centering{212213}&\centering{381}&\centering{14/20}& \centering{11028}& \\
 \centering{GEOM80a}& \centering{360}& \centering{379}&\centering{379}&\centering{380}&\centering{360}&\centering{1510}&\centering{363}& \centering{8583}&\centering{361}&\centering{41235}&\centering{361}&\centering{16/20}& \centering{6103}& \\
 \centering{GEOM80b}& \centering{139}& \centering{145}&\centering{141}&\centering{141}&\centering{139}&\centering{490}&\centering{141}& \centering{1856}&\centering{139}&\centering{255}&\centering{139}&\centering{20/20}& \centering{188}& \\
 \centering{GEOM90}& \centering{330}& \centering{335}&\centering{332}&\centering{339}&\centering{334}&\centering{1810}&\centering{332}& \centering{4160}&\centering{330}&\centering{4022}&\centering{330}&\centering{20/20}& \centering{2917}& \\
 \centering{GEOM90a}& \centering{375}& \centering{382}&\centering{377}&\centering{382}& \centering{377}&\centering{1910}&\centering{382}& \centering{5334}&\centering{375}&\centering{10427}&\centering{375}&\centering{20/20}&\centering{1282}& \\
 \centering{GEOM90b}& \centering{144}& \centering{157}&\centering{157}&\centering{147}&\centering{147}&\centering{590}&\centering{144}& \centering{1750}&\centering{144}&\centering{211366}&\centering{144}&\centering{13/20}& \centering{14648}& \\
 \centering{GEOM100}& \centering{404}& \centering{413}&\centering{404}&\centering{424}&\centering{404}&\centering{2170}&\centering{410}& \centering{3283}&\centering{404}&\centering{40121}&\centering{404}&\centering{12/20}& \centering{16355}& \\
 \centering{GEOM100a}& \centering{437}& \centering{462}&\centering{459}&\centering{461}&\centering{437}&\centering{2500}&\centering{444}& \centering{12526}&\centering{442}&\centering{381}&\centering{442}&\centering{20/20}& \centering{890}& \\
\centering{GEOM100b}& \centering{156}& \centering{172}&\centering{170}&\centering{159}&\centering{159}&\centering{690}&\centering{156}& \centering{3699}&\centering{156}&\centering{213949}&\centering{156}&\centering{3/20}& \centering{86308}& \\
 \centering{GEOM110}& \centering{376}& \centering{389}&\centering{383}&\centering{392}&\centering{378}&\centering{2510}&\centering{383}& \centering{2344}&\centering{381}&\centering{183}&\centering{381}&\centering{20/20}& \centering{452}& \\
 \centering{GEOM110a}& \centering{488}& \centering{501}&\centering{494}&\centering{500}&\centering{490}&\centering{3120}&\centering{490}& \centering{2318}&\centering{488}&\centering{926}&\centering{488}&\centering{20/20}& \centering{1468}& \\
 \centering{GEOM110b}& \centering{204}& \centering{210}&\centering{206}&\centering{208}&\centering{208}&\centering{790}&\centering{206}& \centering{480}&\centering{204}&\centering{944}&\centering{204}&\centering{20/20}& \centering{230}& \\
 \centering{GEOM120}& \centering{396}& \centering{409}&\centering{402}&\centering{417}&\centering{397}&\centering{2730}&\centering{396}& \centering{2867}&\centering{396}&\centering{inf}&\centering{396}&\centering{13/20}& \centering{15341}& \\
 \centering{GEOM120a}& \centering{549}& \centering{564}&\centering{556}&\centering{565}&\centering{549}&\centering{3690}&\centering{559}& \centering{3873}&\centering{554}&\centering{1018}&\centering{554}&\centering{20/20}& \centering{1743}& \\
 \centering{GEOM120b}& \centering{189}& \centering{201}&\centering{199}&\centering{196}&\centering{191}&\centering{910}&\centering{191}& \centering{3292}&\centering{189}&\centering{213989}&\centering{189}&\centering{12/20}& \centering{14371}& \\
 \hline
\end{tabular}
\begin{tablenotes}
\item[1] Note: the best known results ($k^{*}$) are first found in \cite{Chi} for the following 6 instances: GOEM70, GEOM80a, GEOM80b, GEOM100a, GEOM110, GEOM120a. In \cite{Lai}, the reference \cite{Chi} is unfortunately missed.
\end{tablenotes}
\end{threeparttable}
\end{scriptsize}
\end{table}
\renewcommand{\baselinestretch}{1.0}\large\normalsize

Our second experiment aims to compare our PR algorithm with the best performing algorithms in the literature on the set of 33 larger instances transformed from the BMCP problem. In this experiment, in order to make a systematic comparison with the MITS algorithm which is also designed for $k$-BCP, we set $k$ to the best value obtained by the MITS algorithm for each instance and then solve the corresponding $k$-BCP problem. The results of the experiment are summarized in Table \ref{results_bmcp_instances}, together with the results of five reference algorithms, including two hybrid algorithms \cite{Lim1,Lim2}, the FCNS algorithm \cite{Prestwich2}, a stochastic local search method named OF-SW \cite{Chi}, an evolutionary algorithm (EA)\cite{Malaguti}, and the MITS algorithm \cite{Lai}, where the results of reference algorithms are directly extracted from the literature. To report our results, we show the same statistics as in Table \ref{results_bcp_instances}, where ``-'' in columns 7 and 8 means that the associated results are not reported.

Table \ref{results_bmcp_instances} shows that, compared with the first five reference algorithms, our PR algorithm is able to find better results for more than half of the test instances.  In addition, compared with the MITS algorithm which is a single-solution based algorithm, the PR algorithm is faster for 22 out of 33 instances. In particular, the PR algorithm is almost 10 times faster than the MITS algorithm to find a legal $k^{*}$-coloring for several instances, such as GEOM70, GEOM80, GEOM90b, and GEOM120b. Furthermore, for one difficult instance (GEOM120) the MITS algorithm fails to find the best known result even with a very large computational effort and just obtained a $k^{*}$-coloring with objective function value $f=5$, while the PR algorithm obtains a success rate of 13/20 in detecting a legal $k^{*}$-coloring. In short, the PR algorithm outperforms the MITS algorithm on most instances and is highly competitive compared with the best performing algorithms in the literature.

In addition, the success rates of the present algorithm are higher than 12/20 except for three difficult instances (GEOM60b, GEOM70b, and GEOM100b), demonstrating to some extend the robustness of the algorithm on these larger problems.

\subsection{Improved Results on BMCP instances}
\label{further_results}

\renewcommand{\baselinestretch}{1.0}\huge\normalsize
\begin{table}\centering
\caption{Improved Results and Computational Statistics for BMCP instances} \label{further_bcmp_instances}
\begin{scriptsize}
\begin{tabular}{p{1.8cm}|p{1.2cm} p{1.2cm} p{1.2cm} p{1.2cm} p{1.2cm} p{1.2cm} p{0.15cm}}
  \hline
  \centering{{Instance}}&\centering{$k_{best}$}&\centering{$k^{*}$ }&\centering {$k$}&\centering{$suc$}&\centering{$T_{ave}$} &\centering{$k-k^{*}$} & \\
  \hline
  \centering{GEOM50a}&\centering{\textbf{312}}& \centering{314}& \centering{\textbf{312}}&\centering{1/20}&\centering{270860} & \centering{\textbf{-2}}&  \\
  \centering{}&\centering{}& \centering{}& \centering{\textbf{313}}&\centering{12/20}&\centering{14488} & \centering{\textbf{-1}}&  \\
  \centering{GEOM60a}&\centering{\textbf{354}} &\centering{356}& \centering{\textbf{354}}& \centering{7/20}&\centering{34580} & \centering{\textbf{-2}}&\\
  \centering{}& \centering{}&\centering{}& \centering{\textbf{355}}& \centering{14/20}&\centering{13400} & \centering{\textbf{-1}}&\\
  \centering{GEOM70}& \centering{\textbf{266}}&\centering{267}& \centering{\textbf{266}} & \centering{2/20}&\centering{130844} & \centering{\textbf{-1}}&\\
  \centering{GEOM70a}&\centering{\textbf{466}} &\centering{467}& \centering{\textbf{466}}&  \centering{18/20}&\centering{6952} & \centering{\textbf{-1}}&\\
  \centering{GEOM80}& \centering{\textbf{380}}&\centering{381}& \centering{\textbf{380}}&  \centering{7/20}&\centering{34493} & \centering{\textbf{-1}}&\\
  \centering{GEOM80a}& \centering{\textbf{358}}&\centering{360}& \centering{\textbf{358}}&  \centering{6/20}&\centering{41772} & \centering{\textbf{-2}}&\\
  \centering{}& \centering{}&\centering{}& \centering{\textbf{359}}&  \centering{7/20}&\centering{32911} & \centering{\textbf{-1}}&\\
  \centering{GEOM80b}&\centering{\textbf{138}}& \centering{139}& \centering{\textbf{138}}&  \centering{20/20}&\centering{705} & \centering{\textbf{-1}}&\\
  \centering{GEOM90}& \centering{\textbf{328}}&\centering{330}& \centering{\textbf{328}}&   \centering{2/20}&\centering{134941} & \centering{\textbf{-2}}&\\
  \centering{}& \centering{}&\centering{}& \centering{\textbf{329}}&   \centering{12/20}&\centering{15151} & \centering{\textbf{-1}}&\\
  \centering{GEOM90a}& \centering{\textbf{372}}&\centering{375}& \centering{\textbf{372}}&  \centering{1/20}&\centering{282456} & \centering{\textbf{-3}}&\\
  \centering{}& \centering{}&\centering{}& \centering{\textbf{373}}&  \centering{7/20}&\centering{32813} & \centering{\textbf{-2}}&\\
  \centering{}& \centering{}&\centering{}& \centering{\textbf{374}}&  \centering{17/20}&\centering{6027} & \centering{\textbf{-1}}&\\
  \centering{GEOM100a}&\centering{\textbf{436}}& \centering{437}& \centering{\textbf{436}}& \centering{16/20}&\centering{9108}  & \centering{\textbf{-1}}&\\
  \centering{GEOM110}& \centering{\textbf{375}}&\centering{376}& \centering{\textbf{375}}&  \centering{9/20}&\centering{25401}  & \centering{\textbf{-1}}&\\
  \centering{GEOM110a}& \centering{\textbf{482}}&\centering{488}& \centering{\textbf{482}}&  \centering{17/20}&\centering{9819} & \centering{\textbf{-6}}&\\
  \centering{}& \centering{}&\centering{}& \centering{\textbf{483}}&  \centering{18/20}&\centering{6454} & \centering{\textbf{-5}}&\\
  \centering{}& \centering{}&\centering{}& \centering{\textbf{484}}&  \centering{20/20}&\centering{3755} & \centering{\textbf{-4}}&\\
  \centering{GEOM110b}& \centering{\textbf{201}}&\centering{204}& \centering{\textbf{201}}&  \centering{5/20}&\centering{47653} &\centering{\textbf{-3}}&\\
  \centering{}& \centering{}&\centering{}& \centering{\textbf{202}}&  \centering{19/20}&\centering{5580} &\centering{\textbf{-2}}&\\
  \centering{}& \centering{}&\centering{}& \centering{\textbf{203}}&  \centering{20/20}&\centering{1017} &\centering{\textbf{-1}}&\\
  \centering{GEOM120a}& \centering{\textbf{539}}&\centering{549}& \centering{\textbf{539}}&  \centering{6/20}&\centering{45147} &\centering{\textbf{-10}}&\\
  \centering{}& \centering{}&\centering{}& \centering{\textbf{540}}&  \centering{6/20}&\centering{45369} &\centering{\textbf{-9}}&\\
  \centering{}&\centering{}& \centering{}& \centering{\textbf{541}}&  \centering{9/20}&\centering{28124} &\centering{\textbf{-8}}&\\
 \hline
\end{tabular}
\end{scriptsize}
\end{table}
\renewcommand{\baselinestretch}{1.0}\large\normalsize

In order to check whether the results in Table \ref{results_bmcp_instances} can be further improved, we carry out the following additional experiment. First, we determine the smallest $k$ for which a legal $k$-coloring can be reached as follows: (1) The value of $k$ was first set to the best known one and then our PR algorithm searches for a legal $k$-coloring. (2) Once a legal $k$-coloring is found, we set $k$ to $k-1$ and run our PR algorithm again. (3) Step (2) is repeated until the value of $k$ cannot be further improved during 10 consecutive runs of the PR algorithm, and the best result found is recorded as $k_{best}$.

Second, to assess the performance of the PR algorithm for different values of $k$ on the improved instances, we run the PR algorithm independently 20 times for each $k$ from $k_{best}$ to $k_{best}+m$ (where $k_{best}+m < k^{*}$ and $0\leq m < 3$). The computational results are summarized in Table \ref{further_bcmp_instances} with the same statistics as in Table \ref{results_bmcp_instances}. The stopping condition is identical to that used in  section \ref{bmcp} for each run of the PR algorithm.

These tests show that our PR algorithm is able to improve the best known results listed in Table \ref{results_bmcp_instances} for 14 out of 33 instances, and the improvement is impressive for some instances, such as GEOM120a which is solved by using 10 fewer colors than the current best solution. Tables  \ref{results_bmcp_instances} and  \ref{further_bcmp_instances} disclose that our PR algorithm improved or matched the previous best known results for all the BMCP instances.

On the other hand, one can observe that for most instances, a smaller value of $k$ usually corresponds to a lower success rate and a longer average computing time for detecting a legal $k$-coloring. This can be explained by the fact that finding a legal $k$-coloring is much more difficult when $k$ is smaller than the best known value. However, the average computing time is still acceptable for most instances.

\section{Analysis and Discussions }
\label{discussion}

We now turn our attention to analyzing and discussing several important features of our algorithm, including the TS algorithm and the relinking procedure.

\subsection{Importance of the TS Algorithm in the Proposed PR Algorithm}

\begin{figure*}[!t]
\centering
\subfigure[The average objective function value ] {\includegraphics [width=2.6in]{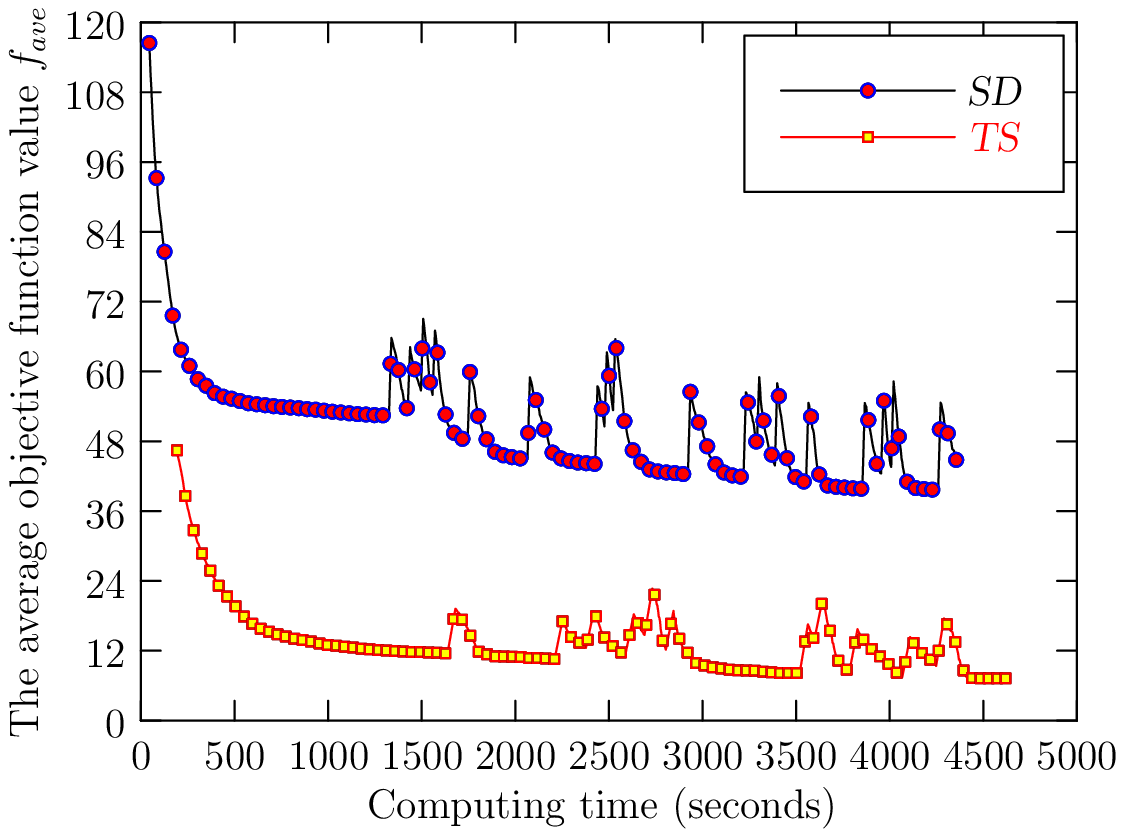}}
\subfigure[The best objective function value] {\includegraphics[width=2.6in]{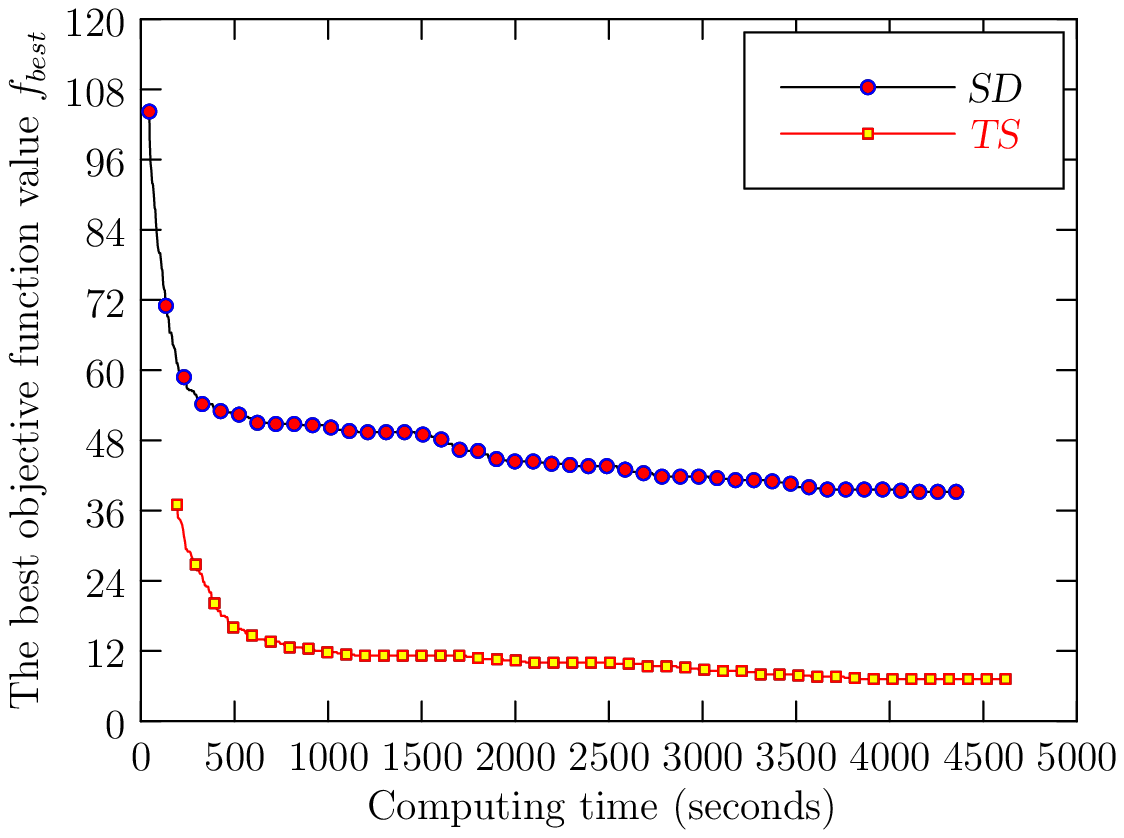}}
\caption{Comparison of TS with SD algorithms within our PR algorithm  }\label{TS_vs_SD}
\end{figure*}

In this section, we assess the role of the TS algorithm within the PR algorithm. For this purpose the following experiment was conducted on a difficult instance named GEOM120 of BMCP with $k=395$.

First, we ran our PR algorithm and separately recorded the best and average objective function values in the population as a function of the computation time. Then, we disabled the TS algorithm and replaced it by the steepest descent (SD) method. We ran this weakened PR algorithm and again recorded the best and average objective function values as a function of computation time. The results are shown in Figure \ref{TS_vs_SD}. Note that all results are based on the average of 5 independent runs.

Figure \ref{TS_vs_SD} demonstrates that using the TS algorithm within the PR algorithm yields much better results than using SD in terms of both the best objective function value and the average objective function value. This experiment highlights the importance of the TS method for the performance of the proposed PR algorithm.

\subsection{ Influence of the Depth $\alpha$ of TS on the Performance }
\begin{figure*}[!t]
\centering
\subfigure[Evolution of the best objective function value ] {\includegraphics [width=2.6in]{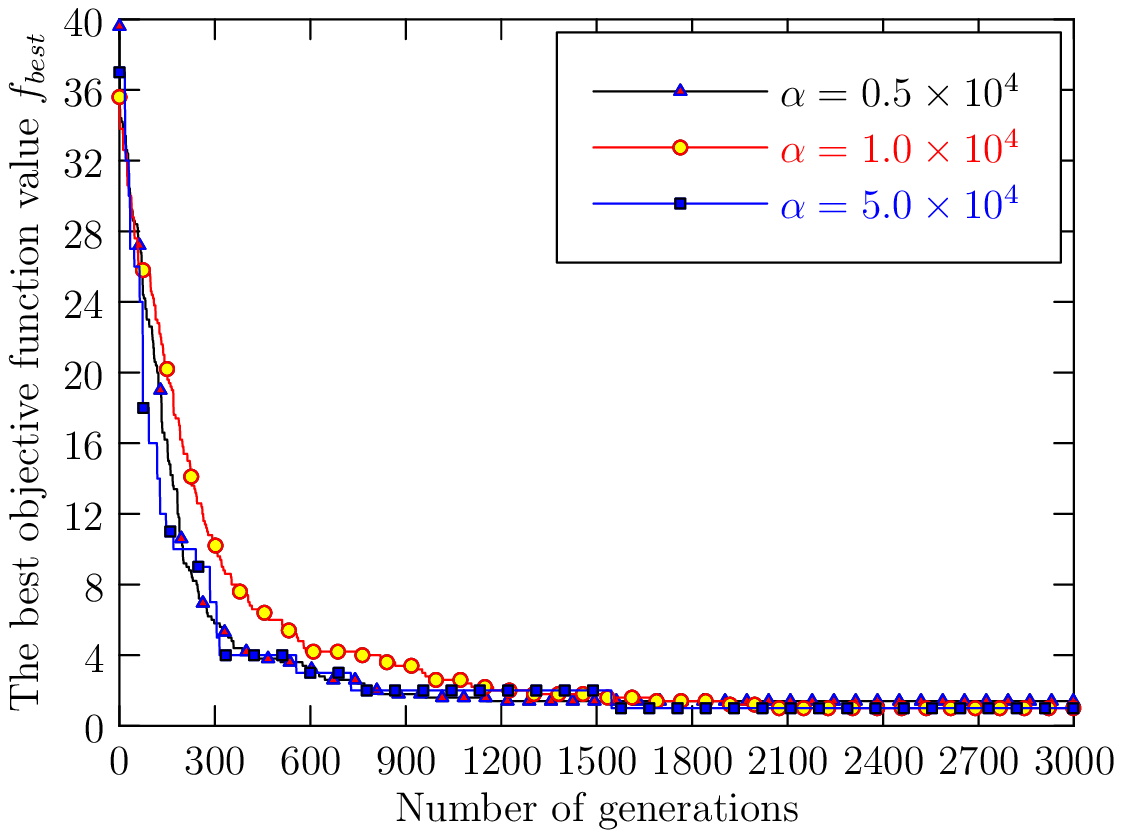}}
\subfigure[Evolution of the computing time ] {\includegraphics[width=2.6in]{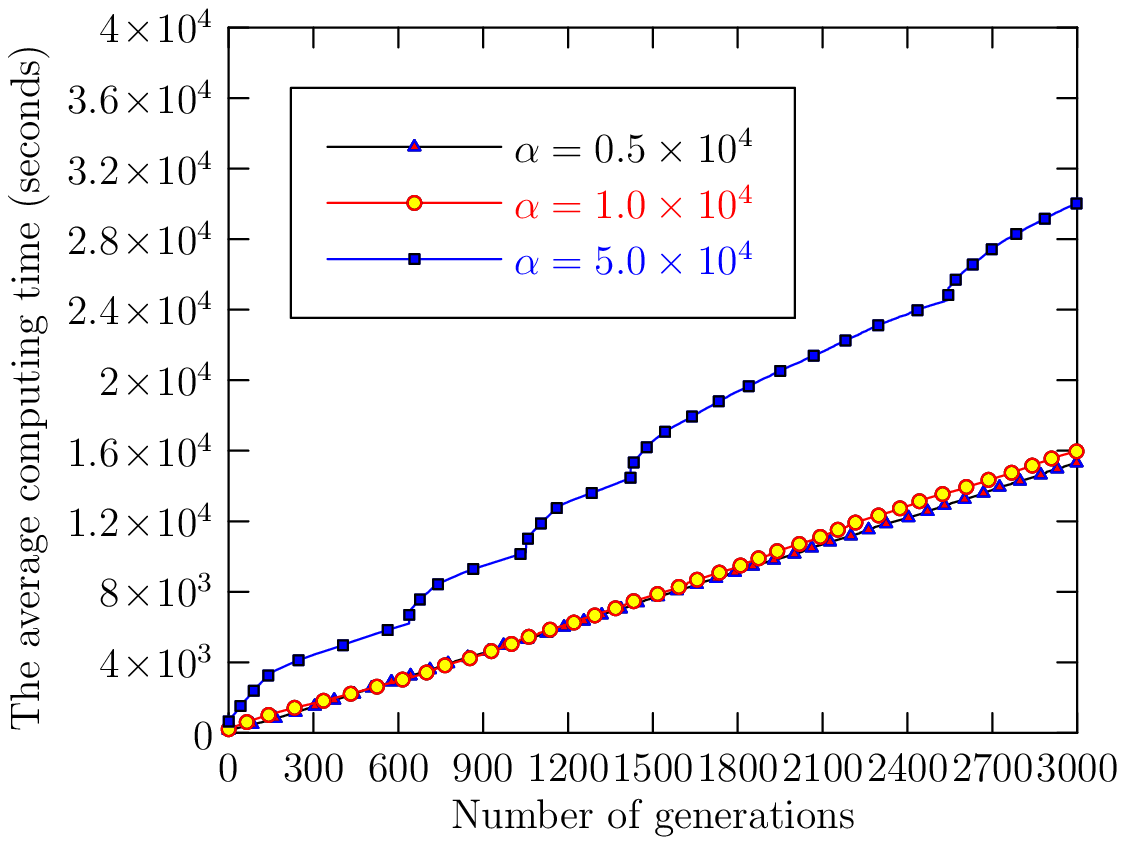}}
\caption{Comparison of the different values for the parameter $\alpha$ } \label{fig_alpha}
\end{figure*}

It is well known that heuristic algorithms usually depend on the settings of parameters used. Therefore, it is meaningful to analyze the influence of the parameters on the performance of the algorithm. In this study, we only discuss one key parameter, i.e., the depth $\alpha$ of TS, since it may normally be expected to have a significant influence on the performance of the PR algorithm.

In most cases, a larger value of $\alpha$ endows the TS method with a stronger local searching ability. However, a larger $\alpha$ also entails a larger computational effort. Therefore, the setting of $\alpha$ is a critical factor in determining a favorable tradeoff between solution speed and quality. To assess the influence of $\alpha$ on the performance of the algorithm, we carried out experiments on the instance GEOM120 of BMCP with $k=395$ using different values of $\alpha$, i.e., $0.5 \times 10^{4}$, $1.0 \times 10^{4}$, and $5.0 \times 10^{4}$, where the stopping condition of the algorithm is the maximum number of generations excluding the population initialization. (We used 3000 generations here.) The evolution of the best objective function value in the population and the computing time with the number of generations are separately plotted in Figure \ref{fig_alpha} for each value of $\alpha$, where the results are based on the average of 5 runs.

One can observe that the PR algorithms with the indicated values of $\alpha$ have quite similar search behaviors, but a larger $\alpha$ value corresponds to a longer computational time. In fact, the PR algorithm with $\alpha = 10^{4}$ and  with $\alpha = 5 \times 10^{4}$ has a success rate of 5/5 for finding the best objective function value ($f=1$). On the other hand, $\alpha = 5 \times 10^{4}$ produces a much longer computing time than $\alpha = 10^{4}$, indicating that a value of $\alpha$ larger than $10^{4}$ will lead to a waste of computational power, and a relatively small value of $\alpha$ like $\alpha= 10^{4}$ suffices to yield a high performance for the present PR algorithm.

\subsection{ Performance Comparison among the  Relinking Procedures }

\begin{figure*}[!t]
\centering
\subfigure[Comparison of the objective function values generated by PR1 and PR2 on the instance  GEOM120 of BMCP ($k=395$) ] {\includegraphics [width=2.6in]{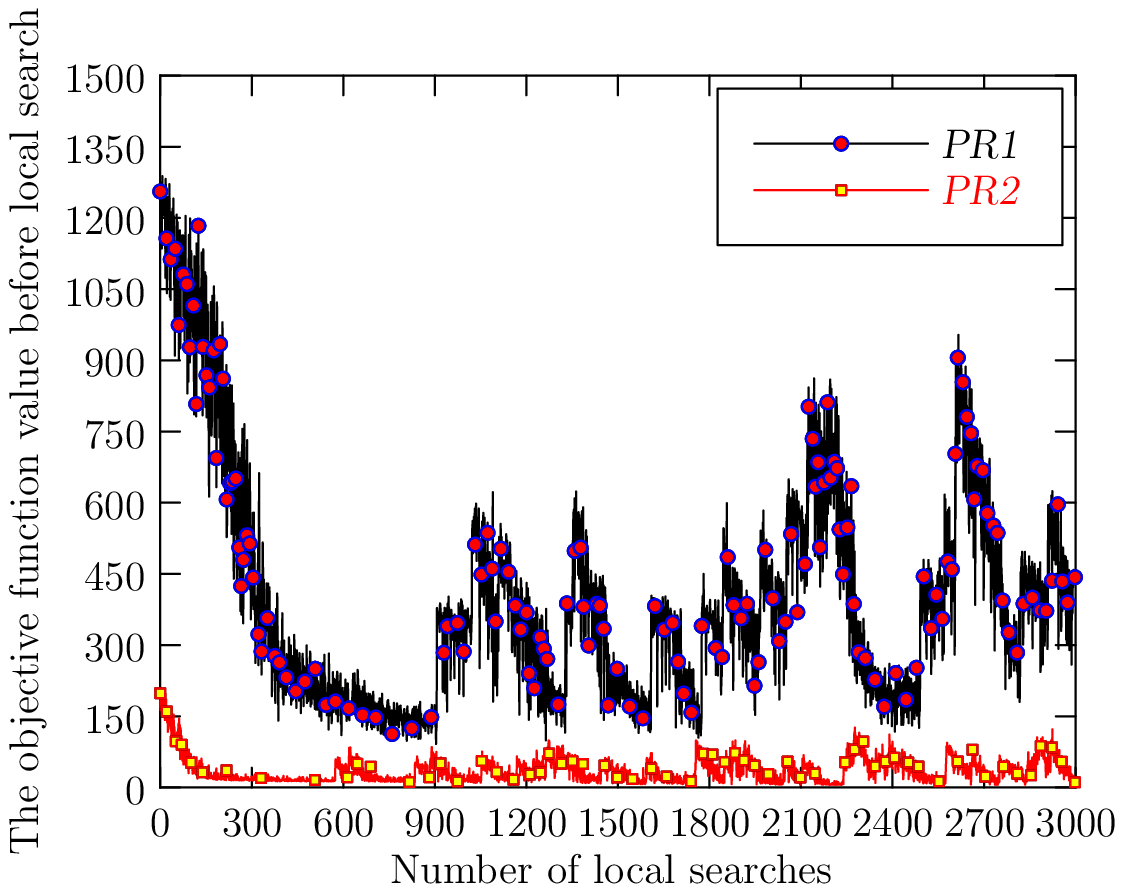}}
\subfigure[Evolution of the best objective function value in the population on the instance GEOM120 of BMCP] {\includegraphics[width=2.6in]{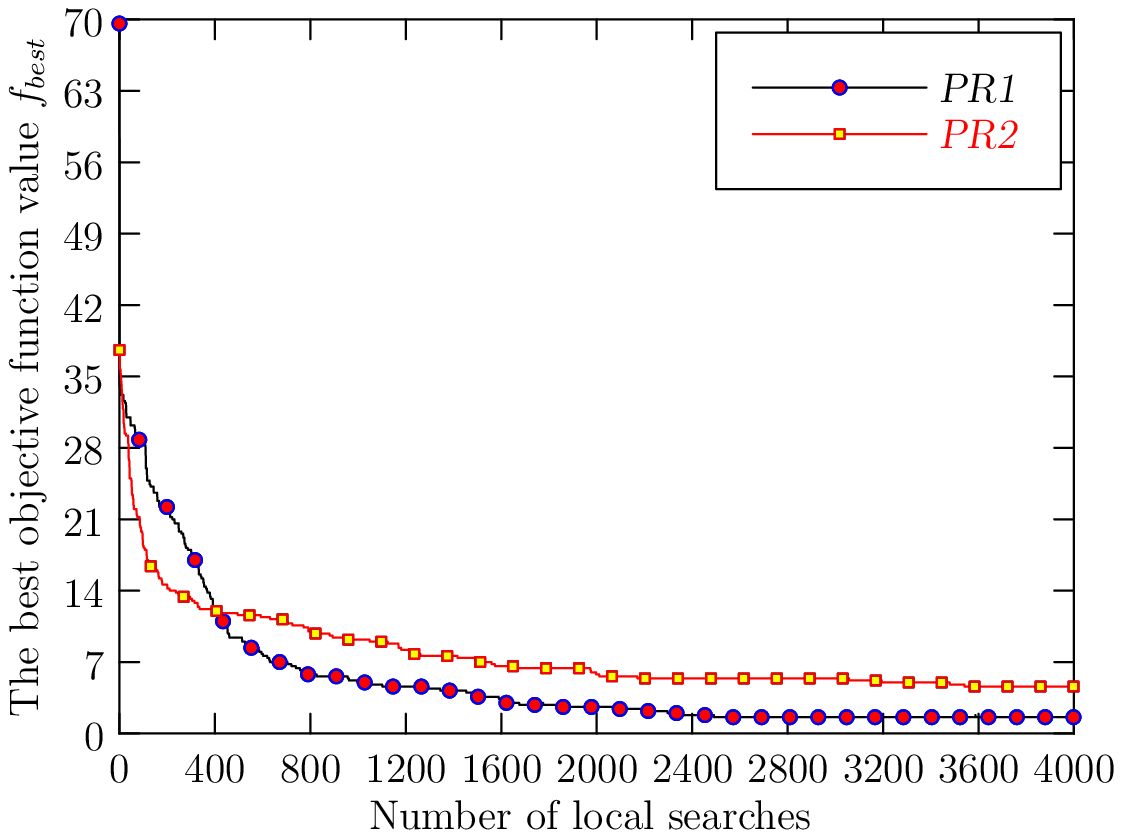}}
\subfigure[Comparison of the objective function values generated by PR1 and PR2 on the instance  GEOM70 of BMCP ($k=265$) ] {\includegraphics [width=2.6in]{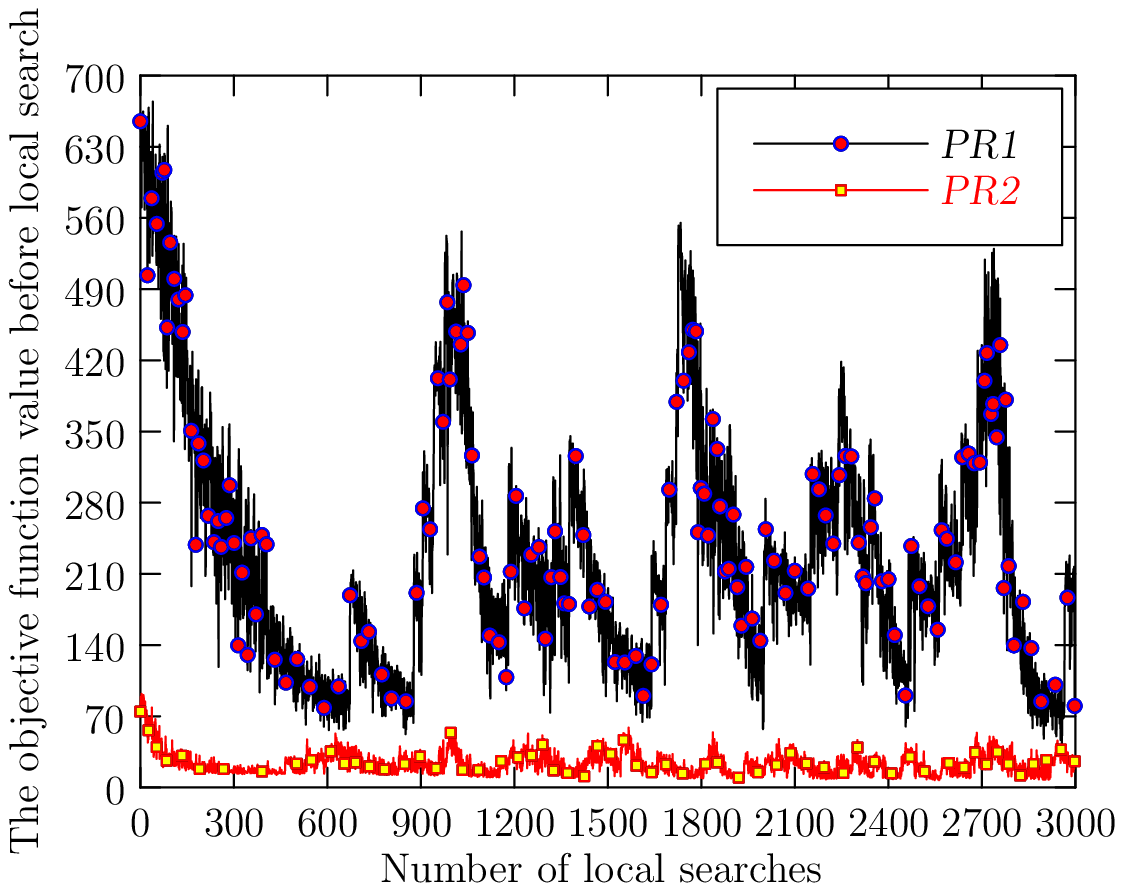}}
\subfigure[Evolution of the best objective function value in the population on the instance  GEOM70 of BMCP ] {\includegraphics[width=2.6in]{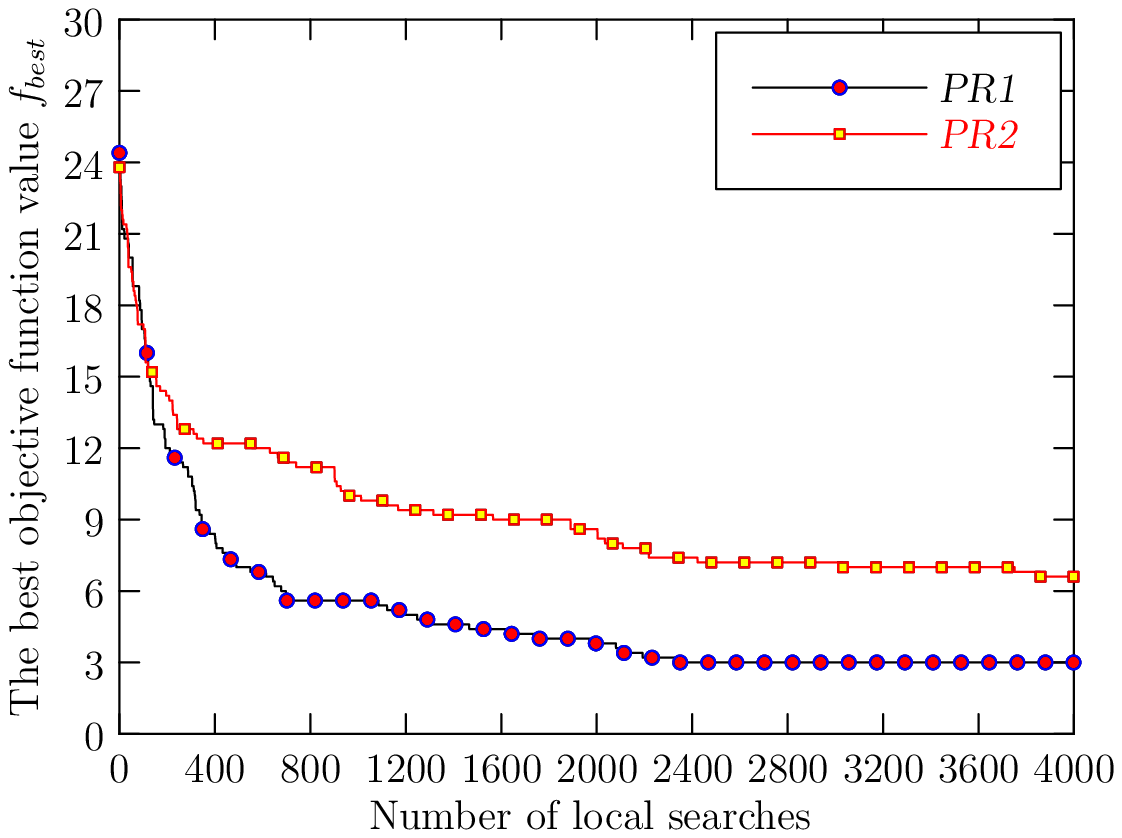}}
\caption{Comparison of PR1 with PR2 relinking procedures }\label{PR1_vs_PR2}
\end{figure*}

As mentioned in \cite{Dorne} and \cite{Ribei}, there are several relinking methods that can be used to generate new offspring solutions, including deterministic methods and randomized methods. PR1 and PR2 used in this study are relatively easy to implement and respectively belong to the classes of the randomized methods and the deterministic methods.

In order to verify whether the relinking procedure plays an important role for the overall performance of our PR algorithm, we compare the PR1 and PR2 relinking procedures with additional experiments. we carry out the experiments using the proposed algorithms respectively with the PR1 or PR2 relinking procedure, based on two difficult instances, i.e., GEOM70 and GEOM120 of BMCP, and all computational results are recorded based on the average of 5 runs of the corresponding PR algorithm.  The objective function values of solutions generated by PR1 and PR2 are compared for the two instances in the sub-figures (a) and (c) of Figure \ref{PR1_vs_PR2},  and the evolutions of the best objective function values in the population with the number of local searches are also plotted in the sub-figures (b) and (d) of Figure \ref{PR1_vs_PR2}.

One can observe from the figures that for the tested instances the greedy PR2 method is able to produce very good offspring solutions in terms of objective function value compared with PR1. However, the PR algorithm with PR1 (randomized relinking method) yields a better performance than the algorithm with PR2 (deterministic relinking method). Therefore, it can be considered that the randomized relinking method is more appropriate than the deterministic relinking method for the tested instances. Intuitively, the search with the greedy relinking method combining with the greedy population updating strategy has troubles to maintain a desirable population diversity as the search progresses. On the other hand, the PR with the randomized relinking method reaches a better balance between diversification and intensification, thus resulting in a stronger searching ability.


\section{Conclusions }
\label{Discussion}
In this paper, we presented a PR algorithm for solving the bandwidth coloring problem and the bandwidth multicoloring problem by incorporating a Tabu Search algorithm with a path relinking procedure. We tested the proposed algorithm on 66 benchmark instances commonly used in the literature. Computational results show that our algorithm is highly competitive in comparison with the best performing algorithms in the literature. In particular, it improved best known results for 15 out of 66 instances and the improvement is very significant for several BMCP cases, yielding solutions with up to 10 fewer colors.

We studied some essential ingredients of the proposed algorithm which shed light on the following points. First, the TS procedure is particularly appropriate as a local optimization method for our PR algorithm. Second, the random relinking procedure (PR1) method is generally more effective than the greedy relinking procedure (PR2) within the present PR framework for the BCP and BMCP problems. Third, the search ability of the present PR algorithm is not very sensitive to the depth $\alpha$ of the tabu search procedure.

\section*{Acknowledgments} This work was partially supported by the National Natural Science Foundation of China (Grant Nos. 61100144 and 61370183) and the Program for New Century Excellent Talents in University of China(NCET 2013).

\bibliographystyle{}

\begin{thebibliography}{}
\bibitem{Chi} Chiarandini M., St\"{u}tzle T., 2007. Stochastic local search algorithms for graph set T-colouring and frequency assignment. Constraints 12(3), 317-403.
\bibitem{Costa} Costa D., 1993. On the use of some known methods for T-coloring of graphs. Annals of Operations Research 41(4), 343-358.
\bibitem{Dorne} Dorne R., Hao, J.K., 1998. Tabu search for graph coloring, T-colorings and set T-colorings. Meta-heuristics: Advances and Trends in Local Search Paradigms for Optimization, Chapter 6, pp77-92, S. Voss, S. Martello, I.H. Osman, C. Roucairol (Eds.), Kluwer Academic Publishers.
\bibitem{fap} Eisenbl\"{a}tter A., Koster A., Frequency assignment websit, 2013.  http://fap.zib.de/index.php.
\bibitem{Ghosh} Ghosh S.C., Sinha B.P., Das N., 2006. Coalesced cap: an improved technique for frequency assignment in cellular networks.  IEEE Transactions On Vehicular Technology 55(2), 640-653.
\bibitem{Glover1} Glover F., 1997. A template for scatter search and path relinking, Lecture Notes in  Computer Science, 1363, 1-51.
\bibitem{Glover2} Glover F., Laguna, M., Mart\'{\i}, R.,  2000. Fundamentals of scatter search and path relinking, Control Cybernetics, 39, 653-684.
\bibitem{Hao1} Hao J.K., Dorne R., Galinier, P., 1998. Tabu search for frequency assignment in mobile radio networks. Journal of Heuristics 4(1), 47-62.

\bibitem{Lai} Lai X.J., L\"{u}, Z.P.,2013. Multistart iterated tabu search for bandwidth coloring problem, Computers and Operations Research, 40(5) 1401-1409.
\bibitem{Lim1}Lim A., Zhu Y., Lou, Q., Rodrigues, B., 2005. Heuristic methods for graph coloring problems. In: Proceedings of the 2005 ACM Symposium on Applied Computing,  pp. 933-939, Santa Fe, New Mexico.
\bibitem{Lim2} Lim A., Zhang X., Zhu Y., 2003. A hybrid method for the graph coloring and its related problems. In: Proceedings of MIC2003: The Fifth Metaheuristic International Conference, Kyoto, Japan.
\bibitem{Lu} L\"{u} Z.P., Hao J.K., 2010. A memetic algorithm for graph coloring. European Journal of Operational Research 203(1), 241-250.
\bibitem{Malaguti} Malaguti E., Toth P., 2008. An evolutionary approach for bandwidth multicoloring problems. European Journal of Operational Research 189(3), 638-651.
\bibitem{Marti} Mart\'{\i} R., Gortazar F., Duarte, A., 2010.  Heuristics for bandwidth coloring problem. International Journal of Metaheuristics 1(1),11-29.
\bibitem{Montemanni} Montemanni, R., Moon, J.N.J., Smith, D.H., 2003. An improved tabu search algorithm for the fixed-spectrum frequency-assignment problem.
 IEEE Transactions On Vehicular Technology 52(3), 891-901.
\bibitem{Phan} Phan V., Skiena S., 2002. Coloring graphs with a general heuristic search engine. In: Computational Symposium on Graph coloring and its generalization,  pp. 92-99. New York.
\bibitem{Prestwich1} Prestwich S., 2002. Constrained bandwidth multicoloration neighborhoods. In: Computational Symposium on Graph coloring and its generalization, pp. 126-133. New York.
\bibitem{Prestwich2} Prestwich S., 2008 Generalized graph colouring by a hybrid of local search and constraint programming. Discrete Applied Mathematics 156(2), 148-158.
\bibitem{Ribei} Ribeiro C.C.C., Resende M.G.C, 2012. Path-relinking intensification methods for stochastic local search algorithms. Journal of Heuristics, 18, 193-214.
\bibitem{Reeves}Reeves C.R., Yamada T., 1998. Genetic algorithms, path relinking and the flowshop sequencing problem. Evolutionary Computation, 6(1), 230-234.
\bibitem{Smith} Smith D.H., Hurley S., Thiel S.U., 1998. Improving heuristics for the frequency assignment problem. European Journal of Operational Research 107(1),76-86.
\bibitem{Trick}Trick M.A., 2002. Computational symposium: graph coloring and its generalization. http://mat.gsia.cmu.edu/COLOR03/.
\bibitem{Velanzuela} Velanzuela C., Hurley S., Smith D.H., 1998. A perturbation based genetic algorithm for the minimum span frequency assignment. In: Proceedings of the 5th international conference on parallel problem solving in nature. pp. 907-916, Amsterdam, The Netherlands.

\bibitem{Walser}Walser J.P., 1996. Feasible cellular frequency assignment using constraint programming abstractions. In: Proceeding of the Workshop on Constraint Programming Applications (CP96), Cambridge, MA, USA.
\bibitem{Wang}Wang W., Rushforth C.K., 1996. An adaptive local-search algorithm for the channel-assignment problem (CAP). IEEE Transactions On Vehicular Technology 45(3),459-466.
\bibitem{Wangyang} Wang Y., L\"{u} Z.P., Glover F., Hao J.K., 2012. Path relinking for unconstrained binary quadratic programming. European Journal of Operational Research 223(3), 595-604
\bibitem{Wu} Wu Q.H., Hao J.K., 2012. Coloring large graphs based on independent set extraction. Computers and Operations Research  39(2), 283-290.

\bibitem{Zhang} Zhang G.Q., Lai K.K., 2006. Combining path relinking and genetic algorithms for the multiple-level warehouse layout problem. European Journal of Operational Research, 169 (2), 413-425.
\end{thebibliography}

\end{document}